\documentstyle[11pt,moriond,epsf,psfig,epsfig]{article}

\def\Journal#1#2#3#4{{#1} {\bf #2}, #3 (#4)}

\def\NPB{{\em Nucl. Phys.} B}
\def\PLB{{\em Phys. Lett.}  B}
\def\PRL{\em Phys. Rev. Lett.}
\def\PRD{{\em Phys. Rev.} D}
\def\ZPC{{\em Z. Phys.} C}
\def\JPG{{\em J. Phys.} G}
\def\PR{{\em Phys. Reports}}

\def\be{\begin{equation}}
\def\ee{\end{equation}}
\def\bea{\begin{eqnarray}}
\def\eea{\end{eqnarray}}

\def\gappeq{\mathrel{\rlap {\raise.5ex\hbox{$>$}}
{\lower.5ex\hbox{$\sim$}}}}

\def\lappeq{\mathrel{\rlap{\raise.5ex\hbox{$<$}}
{\lower.5ex\hbox{$\sim$}}}}

\begin{document}

\pagestyle{empty}
\begin{flushright}
{CERN-TH/98-171}\\
BI-TP 98/14
\end{flushright}
\vspace*{5mm}
\begin{center}
{\bf GENERALIZED VECTOR DOMINANCE AND LOW-$x$ PROTON STRUCTURE%
\footnote{Supported by the BMBF, Bonn, Germany, Contract
 05~7BI92P and the EC-network contract CHRX-CT94-0579.}
} \\
\vspace*{1cm} 
{\bf D. Schildknecht}\\
\vspace{0.3cm}
Theoretical Physics Division, CERN \\
CH - 1211 Geneva 23 \\
and\\
Fakult\"at f\"ur Physik, Universit\"at Bielefeld\\
D-33501 Bielefeld\\
\vspace*{2cm}  
{\bf ABSTRACT} \\ \end{center}
\vspace*{5mm}
\noindent
The low-$x$ HERA data on inelastic lepton-proton
scattering are interpreted in terms of Generalized Vector Dominance.

\vspace*{1.5cm} 
\begin{center}
{\it Talk given at the XXXIIIrd Rencontres de Moriond}\\
{\it on QCD and High Energy Hadronic Interactions}\\
{\it Les Arcs, France, 21--28 March 1998}
\end{center}
\vspace*{1.5cm}
\noindent
\vspace*{0.5cm}

\begin{flushleft} CERN-TH/98-171 \\
BI-TP 98/14\\
May 1998
\end{flushleft}
\vfill\eject

\setcounter{page}{1}
\pagestyle{plain}

%
%
%

Looking back with great pleasure to my participation~\cite{aaa} at the VIIIth
Rencontres de Moriond which took place in M\'eribel-les-Allues in 1973, the subtitle
``Generalized Vector Dominance, 25 years later" seems most appropriate for my
present talk.

There are two basic motivations for returning to the subject, an
experimental one (i) and a theoretical one (ii):
\begin{itemize}
\item[(i)] At HERA two interesting experimental results at low $x$ were
established since HERA started operating in 1992:  first of all, the proton
structure function $F_2(x,Q^2)$ rises steeply with decreasing $x \leq 10^{-2}$
and shows a considerable amount of scaling violations~\cite{bb}.  Secondly, when
analysing the final hadronic state, the H1 and ZEUS collaborations found an
appreciable fraction of final states (appproximately 10\% of the total) of a
typically diffractive nature (``large rapidity gap events") with invariant
masses of the diffractively produced hadronic state up to about
30~GeV~\cite{cc}.
\item[(ii)] With respect to DIS at small $x$, a long-standing theoretical
question concerns the role of the variables $x$ and $Q^2$.  This question has
been most succinctly posed and discussed by Sakurai and Bjorken, as recorded in
the Proceedings of the '71 Electron--Photon Symposium' at Cornell
University~\cite{dd}.  It concerns the transition to the hadron-like behaviour of
photoproduction, more generally, whether concepts similar to the ones used in
photoproduction are relevant in the limit of $Q^2 \rightarrow 0$ only, or
rather in the limit of $x \rightarrow 0$ at arbitrarily large fixed values of
$Q^2$.  Within the framework of QCD there is no unique answer to this question
so far.  We may hope that the HERA low-$x$ data in conjunction with theoretical
analyses will resolve this important issue.
\end{itemize}

In a recent paper~\cite{ee} and in the present talk, I take the point of view
that indeed $x$ is the relevant variable, in the sense that $x \lappeq 10^{-2}$
defines the region in which those features of the virtual photoproduction
cross-section, $\sigma_{\gamma^*p}$, that show a close
similarity to real photoproduction and hadron-induced processes
(Generalized Vector Dominance~\cite{ff}) become important.  Work along these lines, accordingly,
is to be considered as an attempt to quantitatively and directly combine the
above-mentioned two experimental observations at HERA (low-$x$ rise of $F_2$
and diffractive production) within a coherent picture.

Qualitatively, the conceptual basis of Generalized Vector Dominance (GVD) is
strongly supported by:
\begin{itemize}
\item[(i)]
The very existence of diffractive production at low values of the scaling
variable, $(x \lappeq 10^{-2}$), and large $Q^2$, established at HERA and
constituting a ``conditio sine qua non" for the GVD picture:  in GVD, the role
of the low-lying vector mesons, $\rho^0,\omega, \phi$ in photoproduction, at
low $x$ and large $Q^2$, is conjectured to  be taken over by the continuum
of more massive vector states seen in $e^+e^-$ annihilation, which accordingly
ought to be produced diffractively in lepton-proton scattering.
\item[(ii)] The strong similarity in shape between a diffractively produced
state of mass $M_x$ and the state produced in $e^+e^-$ annihilation at the
energy $\sqrt{s_{e^+e^-}} = M_x$.  Compare the thrust and sphericity
distributions shown in Fig.~1 (from Ref.~\cite{ggg}) for diffractive production
and $e^+e^-$ annihilation.  In other words, just as the $\rho^0$ meson produced
in photon-proton interactions looks the same, in good approximation, as the one
seen in $e^+e^-$ annihilation, also the heavy-mass continuum diffractively
produced at HERA looks much the same as the one seen in
$e^+e^-$ annihilation.


%
\begin{figure}[ht]
\centerline{\psfig{figure=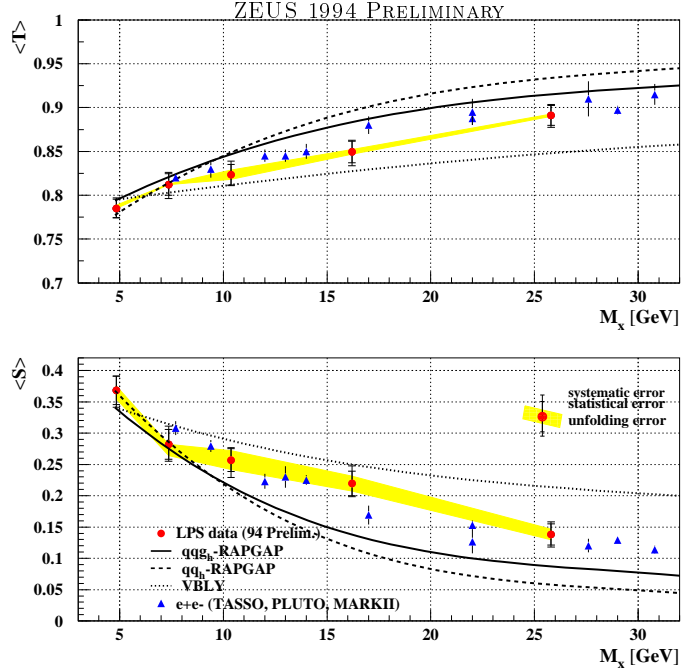,width=10.0cm}}
\caption[]{Thrust $<T>$ and sphericity $<S>$ in diffractive production (ZEUS-LPS data) and
$e^+e^-$ annihilation (from Ref.~\cite{ggg})}
\end{figure}

\item[(iii)] Last, not least, the persistence of shadowing in electron (muon)
scattering from complex nuclei~\cite{hh} at small $x$ and large $Q^2$. 
Diffractive production of high-mass states with photon quantum numbers, the
essential ingredient of GVD, is essential for the destructive interference
responsible for the persistence of shadowing at large values of $Q^2$.
\end{itemize}
Quantitatively, GVD~\cite{ff} starts from a mass dispersion relation which in
general involves off-diagonal transitions in mass and, for the transverse part
of the photon absorption cross-section, takes the form
\begin{equation}
\sigma_T(W^2,Q^2) =
\int dm^2\int dm^{\prime 2}~\frac{\tilde\rho_T(W^2;m^2,m^{\prime 2})m^2m^{\prime 2}}
{(m^2+Q^2)(m^{\prime 2}+Q^2)}
\label{one}
\end{equation}
with appropriate generalization to the longitudinal part $\sigma_L(W^2,Q^2)$
of the total photon absorption cross-section $\sigma_{\gamma^*p}(W^2,Q^2)$. 
While the existence of off-diagonal terms can hardly be disputed from our
experimental knowledge of diffraction dissociation in hadron reactions, and
off-diagonal model calculations have indeed been put forward~\cite{jj}, and
recently reconsidered~\cite{kk}, in applications of GVD, one frequently
approximates (\ref{one}) by an effective representation of diagonal form,
\begin{equation}
\sigma_T(W^2,Q^2) = \int_{m^2_0} \frac{\rho_T(W^2,m^2)m^4}{(m^2 + Q^2)^2}~,
\label{two}
\end{equation}
where the threshold mass, $m_0$, is to be identified with the energy at which the
cross-section for the process of $e^+e^- \rightarrow$ hadrons starts to become
appreciable.  The spectral weight function, $\rho_T(W^2,m^2)$, in (\ref{two})
is proportional to the product of i) the transition strength of a time-like
photon to the hadronic state of mass $m$, as observed in $e^+e^-$ annihilation
at the energy $\sqrt{s_{e^+e^-}} = m$, and ii) the imaginary part of the forward
scattering amplitude of this state of mass $m$ on the nucleon.

For comments on the ansatz for $\sigma_L(W^2,Q^2)$ in the diagonal
approximation,
\begin{equation}
\sigma_L(W^2,Q^2) =
\int_{m^2_0}dm^2~\frac{\rho_T(W^2,m^2)m^4}{(m^2+Q^2)^2}\xi\frac{Q^2}{m^2}~,
\label{three}
\end{equation}
we refer to Refs.~\cite{ff}$^,~$\cite{ee}.  The parameter $\xi$ denotes the ratio
of the longitudinal to the transverse (imaginary) forward-scattering amplitude for
vector states of mass $m$.

When confronting GVD predictions with experimental data, I will discriminate
between an analysis in the region of very small $Q^2$, i.e. $Q^2 \lappeq 1~{\rm GeV}^2$, and
an analysis taking into account the full set of HERA data at low $x$ and values
of $Q^2$ up to the order of $Q^2 \simeq 100~{\rm GeV}^2$.

At small values of $Q^2 \lappeq 1~{\rm GeV}^2$, the dominant contributions to
the integrals in (\ref{two}) and (\ref{three}) stem from low masses, $m^2$, of the 
order of $Q^2$.  In this mass range, the energy dependence for the contributing
hadronic processes may be assumed to be approximately independent of mass $m$,
in generalization of what is known from photoproduction of the low-lying vector
mesons, $\rho^0, \omega$ and $\phi$.  Accordingly, from (\ref{two}) and
(\ref{three}), upon integration, one obtains an expression in which
$W$-dependence and $Q^2$-dependence factorize, the $Q^2$ dependence being
contained in a single pole~\cite{ff} in $Q^2$,
\begin{eqnarray}
\sigma_T(W^2,Q^2)_{\gamma^*p} &=& \frac{m^2_0}{Q^2+m^2_0}~\sigma_{\gamma
p}(W^2)~, \nonumber \\
\sigma_L(W^2,Q^2)_{\gamma^*p} &=& \xi \left[ \frac{m^2_0}{Q^2} \log \left( 1 +
\frac{Q^2}{m^2_0}\right) - \frac{m^2_0}{Q^2+m^2_0}\right]\sigma_{\gamma
p}(W^2)~.
\label{four}
\end{eqnarray}
The comparison of (\ref{four}) with the ZEUS-BPC experimental data is shown in
Fig.~2~\cite{lll}.

%
\begin{figure}[ht]
\centerline{\psfig{figure=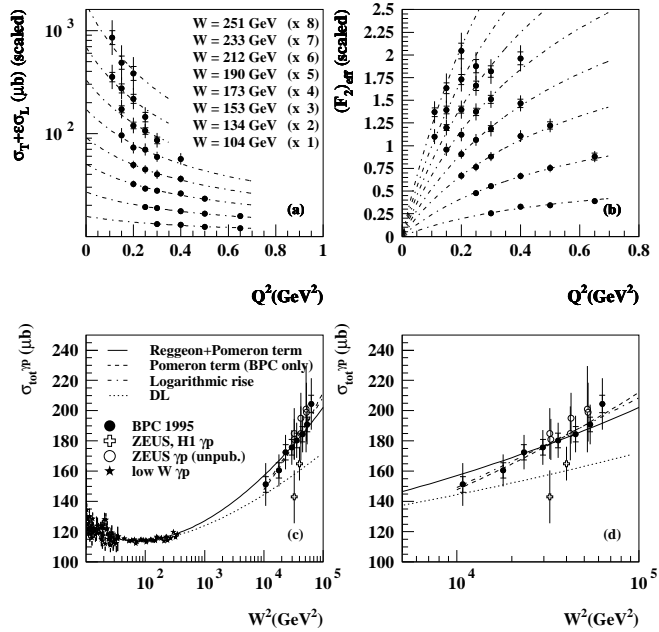,width=10.0cm}}
\caption[]{HERA data  at low $Q^2$ compared with GVD
predictions (from Ref.~\cite{lll})}
\end{figure}

We conclude that
\begin{itemize}
\item[(i)] the $Q^2$ dependence of $\sigma_{\gamma^*p}(W^2,Q^2)$ for $Q^2 \not=
0$ is well described by the GVD ansatz,
\item[(ii)] the extrapolation to $Q^2 = 0$ coincides with (unpublished)
photoproduction results as indicated,
\item[(iii)] the fitted mass scale~\cite{lll} $m_0$ in (\ref{four}),
\begin{equation}
m^2_0 = 0.48 \pm 0.08~{\rm GeV}^2~,
\label{five}
\end{equation}
is reasonable for the threshold energy of $e^+e^-$ annihilation into hadrons,
effectively described by a continuum
starting at $\sqrt{s_{e^+e^-}} = m_0$.
\end{itemize}

Figure 2 also shows a plot of the structure function $F_2(W^2,Q^2)$,
\begin{equation}
F_2(W^2,Q^2) \simeq \frac{Q^2}{4\pi^2\alpha}~\sigma_{\gamma^*p}(W^2,Q^2)~.
\label{six}
\end{equation}
It is amusing to compare this plot of the truly high-energy ZEUS data with a
very similarly-looking plot from 1976 by Robin Devenish (at that time a theorist at DESY)
and myself~\cite{mm} that is based on the ``low-energy" data then
available from the SLAC-MIT collaboration.  The theoretical curves in Fig.~3 are based on
(\ref{four}), using
a fixed input for the threshold mass
\begin{equation}
m^2_0 = 0.36~{\rm GeV}^2~
\label{seven}
\end{equation}
based on theoretical arguments within the off-diagonal ansatz~\cite{jj}.  The
small difference between (\ref{five}) and (\ref{seven}) should not be
overinterpreted, but it is in the right direction, taking into account the fact that
the energy of the SLAC-MIT experiment hardly reaches the charm-production
threshold.


%
\begin{figure}[ht]
\centerline{\psfig{figure=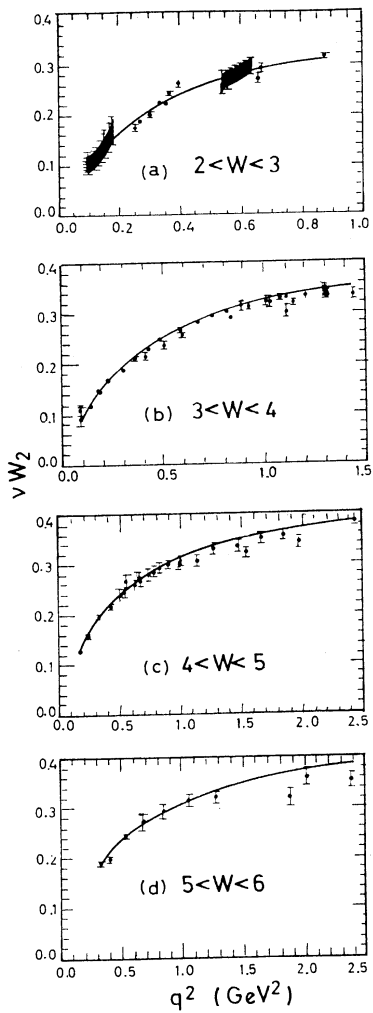,width=5.0cm}}
\caption[]{SLAC-MIT data in comparison with GVD predictions (from Ref.~\cite{mm})}
\end{figure}
At large values of $Q^2$, the simple factorization of the $Q^2$- and
$W^2$-dependence in (\ref{four}) breaks down.  The results in Fig.~4, obtained
by Spiesberger and myself~\footnote{Compare also Ref.~\cite{nn} for an
analysis of the data that is similar in spirit but different in detail.}, are based on a simple logarithmic ansatz for the
spectral weight functions in (\ref{two}) and (\ref{three}),

%
\begin{figure}[ht]
\centerline{\psfig{figure=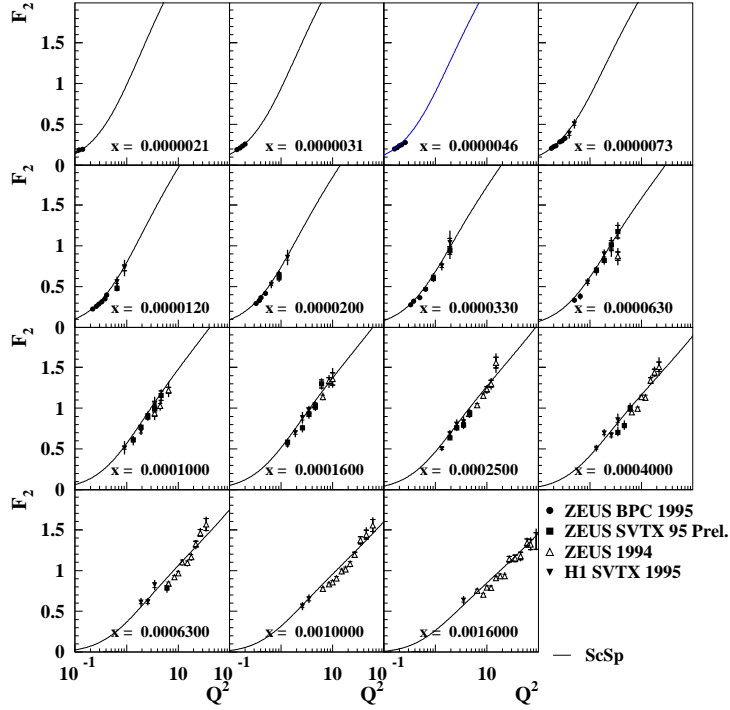,width=10.0cm}}
\caption[]{ZEUS and H1 data compared with GVD predictions (from Ref.~\cite{ee})}
\end{figure}
\begin{equation}
\rho_T(W^2,m^2) = N~\frac{\ln(W^2/am^2)}{m^4}~,
\label{eight}
\end{equation}
which for the photoproduction limit implies a logarithmic rise with energy,
\begin{equation}
\sigma_T(W^2,Q^2 \rightarrow 0) = \sigma_{\gamma p}(W^2) = \frac{N}{m^2_0}
\left(\log \frac{W^2}{a m^2_0} - 1 \right)~,
\label{nine}
\end{equation}
only valid in the truly high-energy HERA regime of $W \gappeq 50$~GeV.  For
details, I refer to Ref.~\cite{ee}.  Let me note, however, that in principle
the magnitude and energy dependence of the photoproduction cross-section is
sufficient to determine the parameters $N$ and $a$, once $m^2_0$ is fixed by
the threshold for the effective $e^+e^-$ continuum.  From the fit to H1 and
ZEUS data we obtained
\begin{eqnarray}
N &=& 5.13 \cdot 4\pi^2\alpha = 1.48~, \nonumber \\
a &=& 15.1~,
\label{ten}
\end{eqnarray}
and for the  parameter $\xi$,

\begin{equation}
\xi = 0.171~.
\label{eleven}
\end{equation}
A brief comment concerns the threshold mass,
\begin{equation}
m^2_0 = 0.89~{\rm GeV}^2~,
\label{twelve}
\end{equation}
obtained in the fit.  Taking into account the fact that the mass dispersion relations
 (\ref{two}) and(\ref{three}) contain  a single threshold mass,
$m_0$, for the effective $e^+e^-$ annihilation continuum, rather than an extra threshold, discriminating the
charm, $c \bar c$, continuum from the rest, a value for $m^2_0$ larger than $m^2_\rho =
0.59~{\rm GeV}^2$, such as (\ref{twelve}), is to be expected.  In fact,
restricting the data set being fitted to a value of $Q^2 \lappeq Q^2_{\rm max}$
with $Q^2_{\rm max} \lappeq 1~{\rm GeV}^2$, thus suppressing the charm contribution, leads to values of $m^2_0$
consistent with (\ref{five}), (\ref{seven}), while for any choice  of $Q^2_{\rm
max} \gappeq 10~{\rm GeV}^2$ a stable value consistent with (\ref{twelve}) is
obtained.  A more detailed analysis  will obviously have to introduce a  separate
threshold mass for the charm, $c \bar c$, continuum in the mass dispersion
relations.

In conclusion, Generalized Vector Dominance provides a unified representation of photoproduction and the low-$x$
 proton
structure in the kinematic range accessible to HERA.  Various refinements remain to be worked out
in the near future, such as a more precise treatment of the charm contribution, the
incorporation of  data at lower energies, and a theoretical analysis of the
diffractively produced final state.  While details are subject to improvement and change,
the principal dynamical ansatz, relating $\sigma_{\gamma^* p}$ or, equivalently, $F_2$ at
low values of $x$ to diffractive scattering (via unitarity) of the states produced in
$e^+e^-$ annihilation, is likely to stand the test of time.

\section*{Acknowledgements}
It is a pleasure to thank Bruce Mellado and Bernd Surrow for useful discussions.


\section*{References}
\begin{thebibliography}{99}
\bibitem{aaa} D. Schildknecht, in Proceedings of the VIIIth Rencontres de Moriond,
ed. Tran Thanh Van, Orsay, France, 1973, Vol. 1, p. 181.
\bibitem{bb} S. Aid {\it et al.}, H1 Collaboration,
\Journal{\NPB}{470}{3}{1996};\\
C. Adloff {\it et al.}, H1 collaboration, \Journal{\NPB}{497}{3}{1997};\\
M. Derrick {\it et al.}, ZEUS Collaboration, \Journal{\ZPC} {72}{399}{1996};\\
J. Breitweg {\it et al.}, ZEUS Collaboration, \Journal{\PLB}{407}{432}{1997}.
\bibitem{cc} T. Ahmed {\it et al.}, H1 Collaboration,
\Journal{\NPB}{429}{477}{1994};\\
M. Derrick {\it et al.}, ZEUS Collaboration, \Journal{\PLB}{315}{481}{1993}.
\bibitem{dd} J.D. Bjorken, in Proceedings of the 1971 Int. Symposium on Electron
and Photon Interactions at High Energies, ed. B.N. Mistry, Cornell University,
1971, p. 281.
\bibitem{ee} D. Schildknecht and H. Spiesberger, hep-ph/9707447;\\
D. Schildknecht, {\it Acta Phys. Pol.} { B28}, 2453 (1997);\\
D. Schildknecht and H. Spiesberger, {\it Acta. Phys. Pol.} to appear.
\bibitem{ff}J.J. Sakurai and D. Schildknecht, \Journal{\PLB}{40}{121}{1972};\\
B. Gorczyca and D. Schildknecht, \Journal{\PLB}{47}{71}{1973}.
\bibitem{ggg} ZEUS Collaboration results, presented by R. Wichmann at DIS98,
Brussels, April 1998.
\bibitem{hh} J. Achman {\it et al.}, EM Collaboration,
\Journal{\PLB}{202}{603}{1998};\\
M. Arnedo, \Journal{\PR}{240}{301}{1994}.
\bibitem{jj} H. Fraas, B.J. Read and D. Schildknecht,\Journal{\NPB}{86}{346}{1975}.
\bibitem{kk} H. L. Frankfurt, V. Guzey and M. Strikman, hep-ph/9712339.
\bibitem{lll}B. Surrow, DESY-THESIS-1998-004;\\
see also B. Mellado, these proceedings.
\bibitem{mm} R. Devenish and D. Schildknecht, \Journal{\it Phys. Rev.} {D14}{93} {1976}.
\bibitem{nn} G. Shaw, \Journal{\PLB}{318}{221}{1993};\\
P. Moseley and G. Shaw, \Journal{\it Phys. Rev.} {D52}{4941}{1995};\\
G. Kerley and G. Shaw, hep-ph/9707465.
\end{thebibliography}
\end{document}